\documentclass[preprint2,10pt,a4paper]{aastex}

\newcommand{\be}{\begin{equation}} 
\newcommand{\ee}{\end{equation}}
\newcommand{\nn}{\mbox{} \nonumber \\ \mbox{} }
\newcommand{\ba}{\begin{eqnarray}}
\newcommand{\ea}{\end{eqnarray}}
\newcommand{\om}{\omega}
\newcommand{\Alfven}{ Alfv\'{e}n }

\newcommand{\B}{{\bf B}}
\newcommand{\J}{{\bf J}}

\newcommand\etal{\textit{et al.\ }}

\newcommand{\Bf}{{magnetic field\,}}

\usepackage{amsmath}
\usepackage{graphicx}
\usepackage{multicol}
\begin{document}
\title{On the dynamics of mechanical failures in magnetized neutron-star crusts}
\author{Yuri Levin$^1,^2$ and Maxim Lyutikov$^3$}
\affil{$^1$ Monash Center for Astrophysics, Monash University, Clayton, VIC 3800, Australia}
\affil{$^2$ Leiden Observatory, Leiden University, Niels Bohrweg 2, Leiden, the Netherlands}
\affil{$^3$ Department of Physics, Purdue University, 525 Northwestern Avenue, West Lafayette, IN, USA}
\email{yuri.levin@monash.edu.au, lyutikov@purdue.edu}
\begin{abstract} \noindent
We consider the dynamics of a mechanical failure induced by a shear stress  in a strongly magnetized  neutron-star crust.  We show that even if  the elastic properties of the crust allow the creation of a shear crack, the strongly sheared magnetic field around the crack leads to a back-reaction from the Lorentz force which does not allow large relative displacement of the crack surfaces. Instead, the global evolution of the crack proceeds on a slow {resistive time scale}, and is unable to release any substantial mechanical energy.  Our calculations demostrate that for {\it some} magnetic-field configurations, 
the magnetic forces cause, effectively, a plastic deformation of the 
crust when the resulting elastic shear stress 
exceeds the critical  value for
mechanical failure.

%Our calculations demonstrate that for {\it some} magnetic-field configurations,  the magnetic forces cause, effectively, a plastic deformation of the  crust when the resulting elastic shear stress  exceeds the critical  value for mechanical failure. 

\end{abstract}
%\clearpage
%\begin{multicols}{2}
\section{Introduction}
In two common astrophysical circumstances, a crust of a neutron star may be placed under considerable stress. Firstly, fast-spinning young pulsars, 
which are reducing their spin frequencies on a short ($\sim1000$ years) timescale, change the shape of their rotational bulges which deforms their 
crusts. Secondly, in magnetars a $\sim 10^{15}$G magnetic field exerts a strong twisting force on their crusts. In both cases the crust yields, and it is reasonable to assume that some of the yield proceeds via explosive ruptures which release the crustal strain and produce star quakes, in analogy with a similar process in the Earth's crust. 

Neutron-star quakes have attracted significant attention in the neutron-star astrophysics folklore. Firstly, it has been proposed that the starquakes in the spinning-down pulsar may help trigger the sudden spin-frequency
increases known as glitches (see, e.g., Link \& Epstein 1996).
Secondly, it has been argued (Thompson \& Duncan 1995, hereafter
TD95, see also Blaes et al. 1989) that starquakes may be responsible for powering flares in Soft-Gamma Repeaters, a sub-class of magnetars. It is mainly this latter suggestion that motivates the 
present study. Briefly, the physical picture is as follows: 

Magnetar emission (see Wood \& Thompson 2004 for a review) is powered by dissipation of a non-potential (current-carrying) magnetic field (TD95,  Thompson, Lyutikov, \& Kulkarni 2002). The field exerts Lorentz force on the crust, which is balanced by induced elastic stress.  For strong enough magnetic fields, Lorentz force may induce a stress that exceeds the critical stress of the lattice.
This leads to breaking of the crust and a release of the seismic waves.  The seismic waves are coupled to the magnetosphere via strong magnetic field and, after converting into the Alfven waves, deposit their energy in a fireball above the neutron-star surface. 

TD95 did not specify how exactly the cracking would proceed; instead, they assumed that
the mechanical failure would occur within some finite volume and that part of the magnetic/elastic energy contained within this volume would be converted into seismic waves (see discussion in their section 2.2). Molecular-dynamical simulations (Horowitz \& Kadau 2009, hereafter HK09, Chugunov \& Horowitz 2010, hereafter CH10) show that indeed
the crust breaks down suddenly when it is shear-stressed above some critical 
level.

A viable alternative theory for the origin of the SGR flares has also been proposed (Lyutikov 2003, 2006), 
This theory does not invoke the starquakes as the flares' origin. Instead, it proposes that
{\it slow }  untwisting  of the internal magnetic field
leads  to gradual twisting of  magnetospheric  field lines, on time scales much longer than the flare duration.
 In this picture the magnetospheric field evolves slowly, perhaps due to the Hall drift (Goldreich \& Reisenegger 1992,  Pons \& Geppert 2007) or plastic flow (Jones 2003). Eventually, the magnetosphere  reaches a dynamical stability threshold  due to increasing energy associated with
current-carrying magnetic field. Then
{\it sudden relaxation } of the  twist outside the
star and associated dissipation and magnetic topology change
lead to flares,  in  analogy with Solar flares and Coronal Mass Ejections (CMEs).
In this scenario, the flare itself is
produced by a  rapid reconfiguration of
the magnetospheric field, which is assisted by a fast reconnection process, much like what happens
in the better-studied
case of a solar flare. Levin \& van Hoven (2011) have shown that the external magnetic-field reconfiguration
during a giant flare would excite the large-amplitude torsional oscillations of the NS crusts; thus in this picture a flare
would preceed the starquake and not the other way around.

It is therefore imperative to understand in detail the mechanics of a mechanical failure of
the neutron-star crust under the strong shear stress. In terrestrial experience, the sudden mechanical failure in response to the shear stress is common and occurs via a macroscopic shear crack, i.e. the breakdown of mechanical elasticity within a thin layer of material.
The crack's rapid progress is easy to understand (e.g., Kostrov 1964): once it forms, the stress at the cracks edge
is strongly enhanced which facilitates the crack's further propagation along
its plane. Often, the efficient crack formation requires the solid's ability to form a small void at the crack's
location, in order for the slippage to be unhindered. Jones (2003) has previously argued that since in the neutron-star crust the pressure is greater than the shear
modulus by two orders of magnitude, the conventional crack that relies on a formation of the void cannot occur.
However, molecular-dynamical simulations of HK09 and CH10 demonstrate a sudden drop
in the shear stress once the shear of the crust coulomb lattice exceeds a critical value of about $0.1$.
Therefore formation of the void is not necessary for the efficient slippage. The situation here is somewhat analogous to that of the deep-focus earthquakes. There it is 
known that the void-aided shear cracks do not form; instead it is thought that a localized shear-stress-induced phase transition reduces the stress and allows for the rapid slippage of tectonic plates
(Bridgeman 1945). 

Two remarks are in order. Firstly,
molecular-dynamic simulations of HK09 and CH10 have not shown shear cracks. On the contrary, the mechanical failure propagated rapidly through the volume of the simulation domain. However, these simulations may not be capable
of capturing  macroscopic elasto-dynamical effects: the domain-size in these simulations is truly microscopic, of order of 100 inter-atomic spaces and is orders of magnitude smaller than e.g. a mean free path of a phonon.  By contrast, the localization of the shear failure into a plane does not depend on the microscopic
nature of the mechanical failure, but is instead a result of the redistribution of the elastic shear stresses once
a localized failure is initiated.

Secondly,
in the theory of mechanical failure it is conventional to distinguish between the brittle and
ductile cracks (see, e.g., Ashby \& Sammis 1990). The former is thought to be initiated at a single location and
proceeds rapidly from the start. The latter begins with a series of unrelated micro-cracks and initially proceeds slowly. The runaway process is initiated
when sufficient number of micro-cracks merge to form a propagating macro-crack. Some materials under shear stress display plastic flows which
may become unstable and concentrate into so-called shear-bends  (see, e.g., Molinari 1997). 
All of these processes have an important feature in common:
the mechanical failure and the rapid shearing motion that follows is 
concentrated into a narrow layer. This feature plays the central
role in our paper.

In this paper we show that the dynamics of a thin shear crack
would be strongly affected by  the magnetic tension inside the crust of a strongly magnetized neutron star.  
In particular we demonstrate that the energy release from a thin crack would be strongly suppressed, since the magnetic field provides mechanical connection between the two slipping sides of the crack and rapidly suppresses the slippage\footnote{A useful terrestrial analogy is the brittle concrete reinforced with metal
rods against fracture. Magnetic fields play a role of the metal rods.}. 
The energy released from a thin crack falls short  by several orders of magnitude to 
power SGR flares. The strong influence of the magnetic
field on the rupture dynamics has no counterpart in the Earth crust, where magnetic field is relatively small
and its coupling to the mechanical motion is weak due to the crust's small conductivity. We  show
that for some magnetic-field configuration the crust response to the magnetically-induced shear stress 
results in what is effectively a plastic deformation rather than sudden rupture, and we construct an explicit model of such 
deformation.

The plan for this paper is as follows. In sections 2 we describe our dynamical solutions for the
magnetized thin crack, and show how magnetic field suppresses seismic energy release. In section 3 we discuss an explicit example with effectively plastic deformation under the action of the magnetar-strength magnetic field.
In section 4 we conclude.

\section{Dynamics of a thin crack}
%Failures in the shear-stressed materials occur within the thin layers. These shear cracks,
%once they are  locally seeded, produce a strong increase the stress aroung their edges, which
%causes further crystal failure. Thus the shear cracks 
 %propagate\footnote{Neither HK nor CH observe failure by shear cracks in their simulation domain.
%However, one has to keep in mind that these are microscopic compared to the scale of the
%crust. Formation of the shear crack is a feature of global elasticity theory (in particular, of how
%the stress-tensor driven momentum flow is altered when a failure appears), and is not sensitive
%to the microscopic nature of the failure.}, and much work has been done in material scielce and geophysics
%to theoretically study the speed of this propagation. 
In order to estimate the maximal
amount of energy released by a fracture, we assume that  it appears instantly as a planar slab of
infinitesimally small thickness and an infinite lateral extent, and that inside this slab the shear modulus and viscosity
suddenly become zero.  
%\subsection{The set-up and dynamical equations}
We build the simplest-possible model which faithfully represents the essential physical aspects
of the system.
The space is assumed to be filled with the homogeneous material of shear modulus
$\mu$, and the material is sheared in the  $x$ direction; the displacement depends on
$z$ only and coordinate $y$ is redundant (i.e., the problem is 2-d; cf.~e.g.~Kostrov 1966). 
The crack appears at $t=0$ along the plane  $z=0$, thus splitting elastic medium into
lower and upper half-spaces. The vertical magnetic field
component $B_z$ is taken to be homogeneous, 
the horizontal magnetic field component $B_x(z)$ is $x$-independent, and $B_y=0$. The system is assumed to
be initially
in equilibrium,  with
\begin{equation}
\partial (T^{\rm el}_{xz}+T^{\rm mag}_{xz})/\partial z =0
\label{statics}
\end{equation}
where $T^{\rm el}$ and $T^{\rm mag}$ are the elastic and magnetic stress tensors, respectively.
The above condition can be re-written as
\be
\B\times \J + T^{\rm el}_{ik,k} =0
\ee
where 
\be 
T^{\rm el}_{ik} = 2 \mu \xi_{ik}, \,
\xi_{ik} ={1\over 2} \left( \xi_{i,k} +\xi_{k,i} \right)
\ee
\B is the \Bf, \J is the current density,  $\mu$ is the shear modulus (assumed spacially constant),  $\xi_{ik}$ is strain tensor, and $\xi_i$ is the displacement vector.  In the incompressible medium $\xi_{kk}=0$, so that $\sigma_{ik,k} =  \mu \Delta \xi_i$.
 The static  \Bf\  can be expressed in terms of flux function $P(x)$
  \be 
  B_x = B_z \partial_z P(z).
  \ee
  The mechanical equilibrium in the $x$ direction becomes
  \be
 { B_0^2 \over 4 \pi} P^{\prime \prime} = \mu \xi_{0}^{\prime \prime}
 \ee
 (primes denote differentiation with respect to $z$). 
  Thus, the initial displacement is given by
 \be
 \xi_0=  { B_0^2 \over 4 \pi \mu} P+k_1+k_2*z,
 \ee
where $k_1$ and $k_2$ are constants.

The quantities $T^{\rm el}_{xz}$ and $T^{\rm mag}_{xz}$ represent the flow of the $x$ component of momentum density in the
$z$ direction [see, e.g., Landau \& Lifshitz 1956], and thus determine
the dynamics of the system.

When the crack appears at $t=0$ along the plane  $z=0$, thus splitting elastic medium into
lower and upper half-spaces, it hinders the momentum flow since $T^{\rm el}=0$ in the crack\footnote{This
assumption is most favorable for the energy release. One could just as well assume that
$T_{\rm el}$ takes some lower than the initial but non-zero value}. The momenta of opposite sign accumulate at the upper and lower sides of the crack, and
they move in the opposite direction. In the unmagnetized case this motion is indefinite, and 
all of the elastic energy of the medium can be released. However, the magnetic tension in the
$z$ direction results in a rapid change of the magnetic $xz$-stress within the crack and, as we show below,
a rapid re-establishment of the initial momentum flow through the crack. This results in the slipping
motion being stopped and
the associated energy release being suppressed. 

\subsection{Motion generated by the crack}

Let $\xi(z,t)$ be the $x$-displacement from the initial equilibrium, and let $b_x(z,t)=\delta B_x/B_z$,  where
$\delta B_x$ is the change in $B_x$ compared to the initial value. The variables are chosen so that
$\xi(z,0)=\dot{\xi}(z,0)=b_x(z,0)=0$. The force balance in the $x$ direction and the induction equations become
\begin{eqnarray}
\ddot{\xi}&=&c_A^2 b_x^{\prime}+c_{\rm el}^2 \xi^{\prime\prime}\label{dyneq1}\\
\dot{b}_x&=&\dot{\xi}^{\prime}+\eta b_x^{\prime\prime}\label{dyneq2}
\end{eqnarray}
Here $\eta$ is the magnetic diffusivity, and $c_A=B_z/\sqrt{4\pi\rho}$ and $c_{\rm el}=\sqrt{\mu/\rho}$ are the Alfven and the elastic shear velocity, respectively, where
$\mu$ is the shear modulus. Our notation is $\dot{\xi}=\partial\xi/\partial t$, $\xi^{\prime}=\partial\xi/\partial z$, etc.

Eqns (\ref{dyneq1}-\ref{dyneq2}) describe the  behavior of the  elastic-resistive medium. Its normal modes are considered in Appendix \ref{Resistive}. There two types of modes: the slowly-damped mechanical shear waves of the elastic medium with \Bf,  and resistive diffusion-type modes.  Next we demonstrate that a sudden crack in a magnetized elastic medium excites mostly the slow resistive modes, while the amplitude of the magnetized elastic modes is very small.

A convenient way to solve initial-value problem with moving boundaries is to use the Laplace transform. We use the
following notation: the Laplace transform of a function $f(t)$ is given by
\begin{equation}
\hat{f}(p)=\int_0^{\infty} f(t) \exp(-pt).
\end{equation}
The behavior of $f(t)$ at late times $t\gg1$ is determined by the behavior of $\hat{f}(p)$ for small values of $p\ll1$.

We now take Laplace transform of the system of homogeneous equations (\ref{dyneq1}) and (\ref{dyneq2}) and look for solutions of the form $\exp[\lambda(p) z]$. We get the following 
general solution:
\begin{eqnarray}
\hat{\xi}(z,p)&=&\hat{A}(p)\exp(-\lambda_1z)+\hat{\bar{A}}(p)\exp(\lambda_1z)\nonumber\\
                    &  &+\hat{B}(p)\exp(-\lambda_2z)+\hat{\bar{B}}(p)\exp(\lambda_2z),\label{sol}
\end{eqnarray}
where
\begin{equation}
\lambda_1={c_t\over c_{\rm el}}\sqrt{p\over \eta},
\end{equation}
and 
\begin{equation}
\lambda_2=p/c_t
\end{equation}
The dispersion relation is obtained by using the smallness $\eta p/ c_{\rm el}^2\ll 1$. This is an excellent approximation (see below) for the
late-time behaviour of the system. 
%However,
%we will henceforth keep the terms of order $\sqrt{\eta p}/c_{\rm el}$; this will give us some insight into
%early-time behaviour. 
The homogeneous solution with $\lambda_2$ corresponds to the familiar shear wave propagating under the combined action of the elastic and magnetic restoring
forces, while that with $\lambda_1$ corresponds to the "diffusion wave", i.e.~the harmonic perturbation damped by magnetic diffusion in the $z$-direction.

We now match the general solution above to the appearance of a crack at $z=0$, $t=0$.
In the upper half-plane $z>0$ only the outgoing waves exist, so $\hat{\bar{A}}=\hat{\bar{B}}=0$. The values of the
coefficients $\hat{A}(p)$ and $\hat{B}(p)$ is obtained from two boundary conditions at the crack surface.

The first boundary condition is straightforward: the elastic stress at the boundary is zero, so the shear at the boundary
has to be zero and therefore
\begin{equation}
{\xi^{\prime}(0_+,t)}=-{\xi_0^{\prime}(z=0)}.
\label{boundary1}
\end{equation}
Here $\xi_0(z)$ is displacement of the crust from the position of zero elastic stress, before the crack appears. 
In the case that the elastic stress balances out the magnetic one, we have 
\begin{equation}
{\xi_0^{\prime}}=-{B_zB_x\over 4\pi\mu}=-{c_A^2\over c_s^2}{B_x\over B_z},
\label{initialshear}
\end{equation}
where $c_s=\sqrt{\mu/\rho}$ is the velocity of a purely elastic shear wave.
Taking the Laplace transform of Eq.~(\ref{boundary1}), we obtain 
\begin{equation}
{\hat{\xi}^{\prime}(0_+,p)}=-{1\over p}{\xi_0^{\prime}(z=0)}.
\label{boundary10}
\end{equation}
From here onwards we shall write the shorthand of $\xi_0^{\prime}$ for $\xi_0^{\prime}(z=0)$.

The second boundary condition is obtained by considering the singularity at the crack,
$\xi^{\prime}=2\xi(0_+)\delta(z)$. Integrating Eq.~(\ref{dyneq2}) across the boundary, using the
fact that $\dot{b}_x$ is finite, and the symmetry $b_x^{\prime}(z)=-b_x^{\prime}(-z)$ we get 
\begin{equation}
\eta b_x^{\prime}(0_+,t)+\dot{\xi}(0_+,t)=0
\end{equation}

Taking the Laplace transform of this equation, we obtain
\begin{equation}
\eta \hat{b}_x^{\prime}(0_+,p)+p\hat{\xi}(0_+,p)=0.
\label{Laplace1}
\end{equation}
From Eq.~(\ref{dyneq1}) we get
\begin{equation}
\hat{b}_x^{\prime}=(1/c_A^2)[p^2\hat{\xi}-c_{\rm el}^2 \hat{\xi}^{\prime\prime}].
\label{Laplace2}
\end{equation}
Substituting Eq. (\ref{Laplace2}) into Eq. (\ref{Laplace1}) we get at the boundary $z=0_+$:
\begin{equation}
p\left[{p\eta\over c_A^2}+1\right]\hat{\xi}=\eta {c_{s}^2\over c_A^2}\xi^{\prime\prime}.
\label{Laplace3}
\end{equation}
 Substituting Eq.~(\ref{sol})
(with $\bar{A}=\bar{B}=0$) into Eq. (\ref{Laplace3})
we obtain (neglecting $\eta p/c_{\rm el}^2$):
\begin{equation}
\hat{B}={c_{\rm el}^2\over c_A^2}\hat{A}.
\label{result1}
\end{equation}
Substituting this and Eq.~(\ref{sol}) into the first boundary condition in Eq.~(\ref{boundary1}), we get
\begin{equation}
\hat{B}(p)=c_t\xi_0^{\prime}{1\over p^{3/2}\left(\sqrt{p}+\sqrt{p_0}\right)},
\label{B}
\end{equation}
where 
\begin{equation}
p_0={c_t^4c_A^4\over c_{\rm s}^6\eta}
\label{p0}
\end{equation}

\begin{figure}[h!]
\includegraphics[width=.49\linewidth]{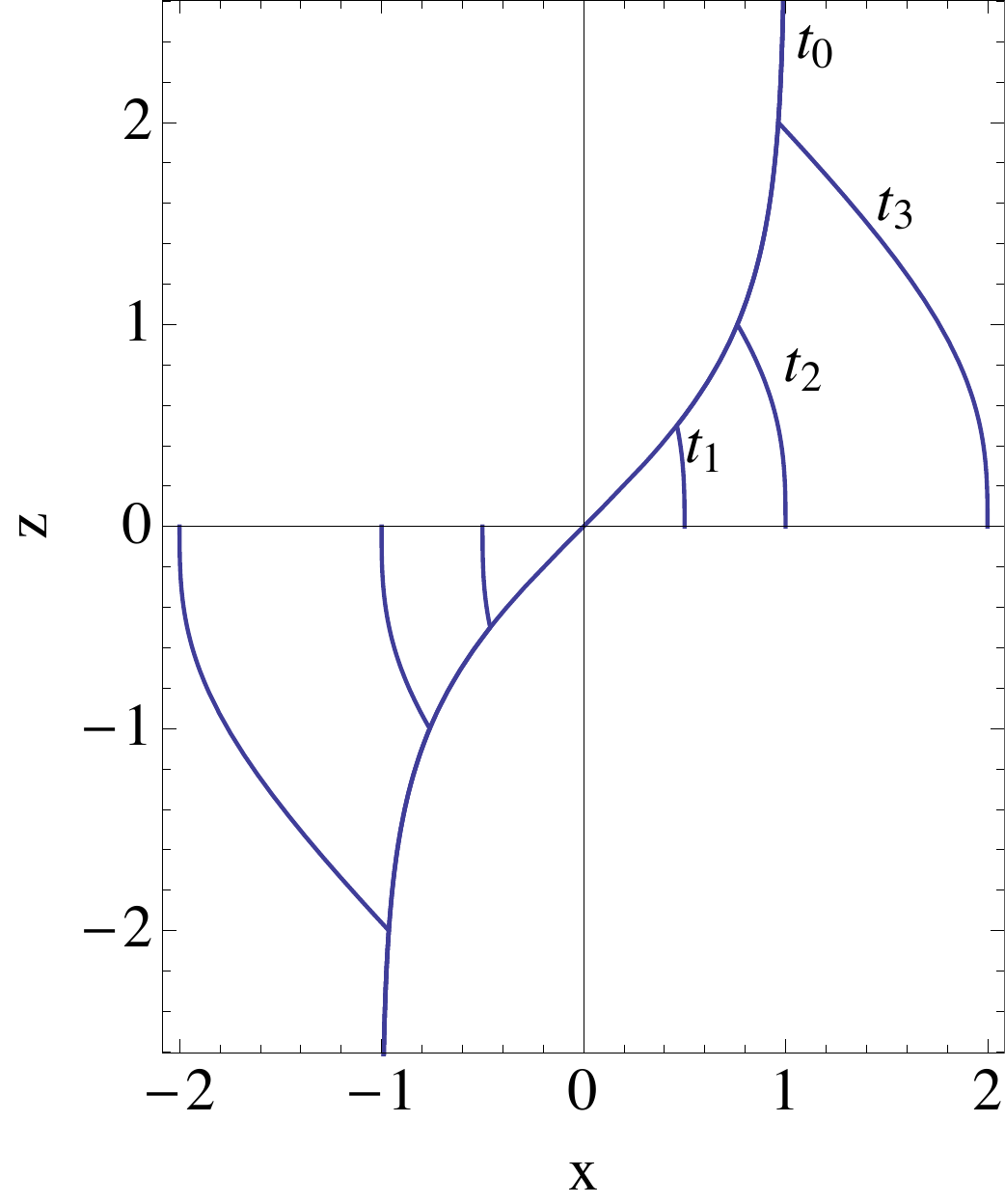}
\includegraphics[width=.49\linewidth]{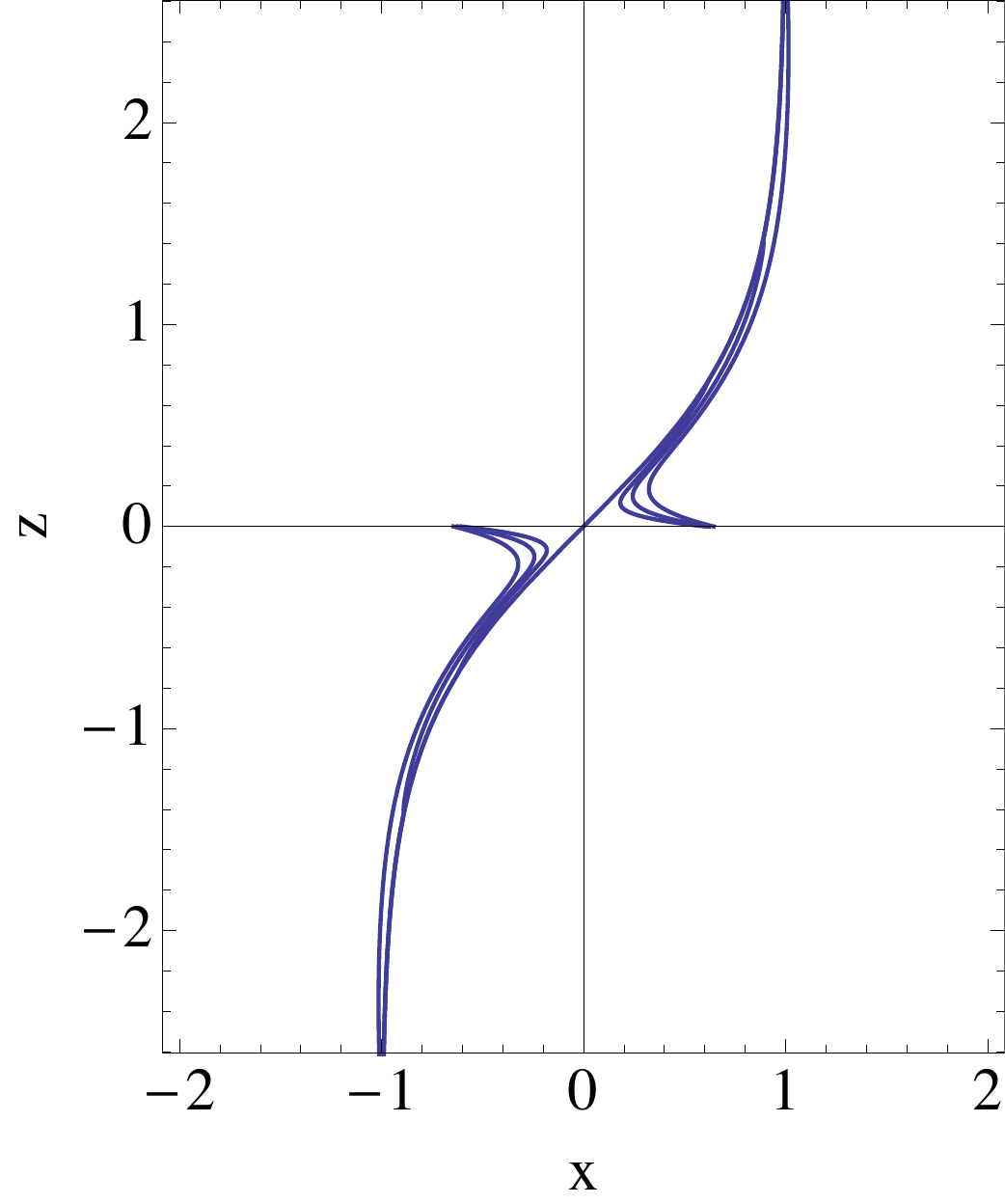}
\caption{ Examples of   the displacement after the crack in a non-magnetized material  ({\it Left Panel}) and magnetized material   ({\it Right Panel})  for different times $t_0 < t_1<t_2<t_3$.  Initial profile is $\xi_0 = \tanh z$, both shear and \Alfven velocities are unity, $\eta = 0.01$.  At $t=t_0$ the crack appears at the $z=0$ plane, launching two shear waves in the non-magnetized case and  two shear   and two resistive waves in the  magnetized case. In the magnetized case the shear waves have very small amplitude, $\propto \sqrt{\eta}$ and do no carry large energy flux. The resistive waves are limited to a narrow region $\Delta z \sim l$.} 
\label{ShearnoB}
\end{figure}

We consider two
useful limiting cases of the above equation (the general solution will be obtained later in this section):\newline
{\it Case (i)}: For unmagnetized crust, $c_A=p_0=0$, and one gets
\begin{eqnarray}
\hat{B}(p)&=&c_{\rm el}\xi_0^{\prime}/p^2,\nonumber\\ 
\hat{A}(p)&=&0.
\end{eqnarray}
The inverse Laplace transform of $\hat{B}(p)$ is given by 
\begin{equation}
B(t)=c_{s}t\xi_0^{\prime}
\end{equation}
and the corresponding displacement in the upper half-plane $z>0$ is given by
\begin{equation}
\xi(z,t)=c_s\xi_0^{\prime}\times\left(t-{z\over c_s}\right)\Theta\left(t-{z\over c_s}\right),
\label{B=0}
\end{equation}
where $\Theta(t)$ is the Heaviside function. This solution
  represents the shear wave launched from the suddenly introduced unmagnetized crack. The velocity jumps discontinuously across the crack,
\begin{equation}
\dot{\xi}(0_+,t)-\dot{\xi}(0_-,t)=2c_{\rm el} \xi_0^{\prime},
\end{equation}
see Fig. \ref{ShearnoB}.

If magnetic field is present, however, this fast slippage is quickly stopped by the rapidly building magnetic tension, and all further slippage is controlled by magnetic-field diffusion. This leads us to consider another 
limiting case. 
\newline

{\it Case (ii)}:  Of major interest for us is the solution at late times, $t\gg 1/p_0$. To these,  only $p\ll p_0$ contribute substantially and in this
limit we have
\begin{equation}
\hat{B}(p)={c_{\rm el}^3\xi_0^{\prime}\sqrt{\eta}\over \sqrt{\pi}c_t c_A^2}p^{-3/2},
\end{equation}
and 
\begin{equation}
\hat{A}(p)={c_A^2\over c_{\rm el}^2} \hat{B}(p)
\end{equation}
Substituting this into Eq.~(\ref{sol}) , we obtain
\begin{equation}
\hat{\xi}(z,p)=\hat{\xi}_B(z,p)+\xi_A(z,p),
\label{2parts}
\end{equation}
where 
\begin{equation}
\hat{\xi}_B(z,p)={c_t\xi_0^{\prime}\over \sqrt{p_0}}{1\over p^{3/2}}\exp\left[{-zp\over c_t}\right],
\label{partB}
\end{equation}
and
\begin{equation}
\hat{\xi}_A(z,p)={c_{\rm el}\over c_t}\xi_0^{\prime}{1\over p^{3/2}}\exp\left[-{c_t\over c_{\rm el}}\sqrt{p\over
\eta}z\right].
\label{partA}
\end{equation}
Taking the inverse Laplace transform, we obtain
\begin{equation}
\xi_B(z,t)=\xi^{\prime}_0{c_{\rm el}^3\over c_A^2 c_t}{2\over\sqrt{\pi}}\sqrt{\eta\cdot(t-z/c_t)}
\Theta(t-z/c_t),
\label{part1}
\end{equation}
\begin{equation}
{\partial \xi_A(z,t)\over \partial z}=-\xi_0^{\prime} \cdot {\rm Erfc}\left[{z\over l}\right],
\end{equation}
and 
\begin{equation}
\xi_A(z,t)=\xi_0^{\prime}l\left[
{\exp\left(-[z/l]^2\right)\over \sqrt{\pi}}-[z/l]\cdot {\rm Erfc}(z/l)\right],
\end{equation}
where 
\begin{eqnarray}
l&=&{2c_{\rm el}\over c_t}\sqrt{\eta t}\nonumber\\
&=&0.02 {c_{\rm el}\over c_t} \left({\eta\over 10^{-5}\hbox{cm}^2 \hbox{s}^{-1}}\right)^{1/2}
\left(t\over 10\hbox{s}\right)^{1/2}\hbox{cm},
\label{ldiffusions}
\end{eqnarray}
(see Fig. \ref{ShearnoB}).

The two contributions have distinct physical character. The $\xi_B(z,t)$ represents a shear wave which is launched
by the slippage at $z=0$ and propagates with the speed $c_t$. It is this contribution which is responsible for the
release of seismic energy into the crust. On the other hand, the $\xi_A(z,t)$ represents the magnetic-diffusion-type
process which operates locally within distance $\sim l$ given by the equation   Eq.~(\ref{ldiffusions}) from the crack.
Within this thin layer the elastic stress of the crust is substantially reduced.

The surface $z=0+$ moves according to
\be
\xi(z=0+) = \xi_0^{\prime}  {c_{\rm el}^2 \over c_A^2 c_t} \sqrt{ 2 \eta t /\pi}
\ee
Thus, within a very short time, 
\be
t \sim  { \xi_0^{\prime} \over 2 \pi}  {c_{\rm el}^2  \eta \over c_A^4} 
\ee
the slippage velocity falls below the elastic shear velocity: \Bf\ effectively stops the relative motion of the crack's surfaces. 

For typical neutron-star crust parameters
(Chamel \& Haensel 2008), one evaluates
\begin{equation}
p_0=10^{21}\hbox{s}^{-1} {\sigma\over 10^{25} \hbox{s}^{-1}}
\left({c_t\over c_{\rm el}}\right)^2\left({c_t\over 10^8 \hbox{cm}/\hbox{s}}\right)^2\beta^2,
\label{omega0num}
\end{equation}
where 
\begin{equation}
\beta=c_A^2/c_{\rm el}^2.
\end{equation}
For magnetar-strength fiels, $\beta\sim 1$ and all times of interest satisfy 
$t\gg 1/p_0$. Therefore the approximations for the case considered here
hold extremely well in a magnetar. The 
 energy flux going out in seismic waves is
\begin{eqnarray}
{d^2E\over dx dy dt}&=&{1\over 2}\rho\dot{\xi}_B(0,t)^2 c_t={1\over \pi\beta^2}{c_{\rm el}^2\over c_t}\rho \left(\xi_0^{\prime}\right)^2 {\eta\over t}.
\label{flux}\\
&=&3\times 10^{14}{\rho\over 10^{14}}{c_{\rm el}^2/c_t\over 10^8}\left({\xi_0^{\prime}\over 0.1}\right)^2{\eta\over 10^{-5}t}
{\hbox{erg}\over \hbox{cm}^2\hbox{s}},\nonumber
\end{eqnarray}
where all the quantities in the expression above are expressed in the cgs units.

Taking an optimistically large crack area of $10^{12}\hbox{cm}^2$, we get the typical
released energy of $\sim 10^{27}$erg for the magnetar-strength field with $\beta\sim 1$.
This is $\sim12$ orders of magnitude short compared to typical weak SGR flares
(Woods \& Thompson 2006). {\it Clearly, thin shear cracks cannot be responsible for the flares}.
We finish this section with 2 remarks.

\paragraph{Mathematics remark.} One can evaluate the inverse Laplace transform of Eq.~(\ref{B}), thus finding the motion due
to appearance of the crack without considering limiting cases.
It is easiest to first evaluate "seismic velocity" $\dot{B}$, whose Laplace transform is $pB(p)$.
The latter could be calculated by noticing that the  inverse Laplace transform
of 
\begin{equation}
\hat{f}(p)={1\over \sqrt{p}(\sqrt{p}+1)}
\end{equation}
is given by
\begin{eqnarray}
f(t)&=&{1\over \pi}\int_{-\infty}^{\infty}{e^{-k^2t}\over k^2+1}dk\nonumber\\
&=&e^t\left[1-\hbox{Erf}\left(\sqrt{t}\right)\right].
\label{ft}
\end{eqnarray}
This can be checked directly from the definition of Laplace transform (the integrals over first $t$
and then $k$ are elementary).
The "seismic velocity"  is then give by
\begin{equation}
\dot{B}(t)=c_t\xi^{\prime}_0 f(p_0t).
\label{Bt}
\end{equation}
It is straightforward to check that the limits $t\ll 1/p_0$ (free slip) and $t\gg 1/p_0$ (magnetic-diffusion-controlled slip) are recovered\footnote{For $x\gg1$, with
high degree of precision
$f(x)= \sqrt{1/(\pi x)}$.}.

\paragraph{Physics remark.} 
We have considered the configuration where there is a substantial magnetic field
in the direction perpendicular to the crack. One could ask whether a crack direction
could adjust itself in such a way that the crack surface would be tangential
to the field lines. From Eq.~(\ref{flux}), we see that the $B_z$ component
would have to be smaller than the average $B$ by a factor greater than $10^3$ in order
for the crack to release sufficient energy for the flare (this is because
$\beta\propto B_z^2$).  It seems very unlikely to us that this type 
of fine-tuned alignment could
be efficient on a large scale, unless there was an elasto-dynamical mechanism that would
drive the crack into alignment with the field. We have so far failed to identify such a mechanism.
%\subsection{The solution}

\section{Plastic flows?}
We now construct an explicit scenario where changing magnetic field configuration induces effectively a
gradual plastic flow in the constant-density crust. We shall assume, as before, $B_z=\hbox{const}$ and consider $B_x$ that is initially increasing with time
\begin{equation}
B_x=Ct[\cosh(z/L)]^{-2}
\label{bxplastic}
\end{equation}
due to some non-MHD process (e.g., Hall drift) the specifics of which are not important for our purpose.
Here $C$ is some constant and $L$ is the vertical length scale. We assume that the $xz$-component of
the magnetic stress  is balanced precisely by the $xz$-component of the elastic stress, i.e.
\begin{equation}
-T^{\rm mag}_{xz}(z)={B_zB_x(z)\over 4\pi}=T^{\rm el}_{xz}(z)=-\mu\xi_0^{\prime}(z).
\label{ineq}
\end{equation}
Let $T^{\rm el}_{\rm crit}$ be the critical shear stress at which the lattice begins to slip. According
to HL09, $T^{\rm el}_{\rm crit}\simeq 0.1\mu$. Once the external magnetic field exceeds the critical
value at height $z$,
\begin{equation}
B_x^{\rm crit}=
{T^{\rm el}_{\rm crit}\over 4\pi B_z},
\label{Bcrit}
\end{equation}
the crust begins to break and forms a shear crack at that location. From Eq.~(\ref{bxplastic}), this
occurs when
\begin{equation}
\cosh(z_{\rm crit}/L)=\sqrt{Ct\over B_x^{\rm crit}}.
\label{zcrit}
\end{equation}
The vertical velocity with which the boundary of the critical domain moves is given by 
\begin{equation}
v_{\rm crit}={dz_{\rm crit}\over dt}={1\over 2}L\sqrt{C\over B_x^{\rm crit} t}[\sinh(z/L)]^{-1}.
\label{vcrit}
\end{equation}
Both the elastic stress and the
$B_x$ are $0$ at the crack's location and are reduced within distance $\delta z\sim \sqrt{\eta\delta t}$, where  $\delta t$ is the time interval that has passed from the crack's formation; see Eq.~(\ref{ldiffusions}). The time interval $\delta t$ between the
formation of two sequential cracks is given by
\begin{equation}
\sqrt{\eta\delta t}\sim v_{\rm crit} \delta t,
\end{equation}
so that
\begin{equation}
\delta t\sim \eta/v_{\rm crit}^2,
\label{deltat}
\end{equation}
and the separation between the cracks is 
\begin{equation}
\delta l\simeq \eta/v_{\rm crit}.
\label{deltal}
\end{equation}
Some numerical estimates are in order. Assuming that the magnetar field evolves due to Hall drift
on a timescale of $\sim 100$yr, one gets
\begin{equation}
v_{\rm crit}\sim {1\hbox{km}\over 100\hbox{yr}}=3\times 10^{-5}\hbox{cm}\hbox{s}^{-1},
\end{equation}
and therefore
\begin{eqnarray}
\delta l&\sim& 0.3\hbox{cm}\nonumber\\
\delta t&\sim& 10^5\hbox{s}\sim\hbox{day}.\label{scales}
\end{eqnarray}
It is on the latter timescale that the magnetic field $B_x$ and the elastic shear $\xi^{\prime}-\xi_0^{\prime}$ relax to zero within the stressed domain where multiple cracks have appeared.

\section{Discussion}
The neutron star crust cracks differently from that of the Earth. In the latter, the quakes are caused
by a sudden elastic stress release along a 2-dimensional surface, the shear crack, which causes slippage
and release of mechanical energy. The magnitude of the release is much greater than the amount
of elastic energy stored within the volume of the crack.  By contrast, in the neutron-star crust magnetic tension
strongly suppresses  the slippage and the energy release.  Infinitely thin cracks release energy 
 via magnetic-field diffusion, too slowly to be able to contribute to the
energetics of magnetar flares.   If  SGR flares
are associated with starquakes which are generated by a crystal failure of the
stressed crust, these constraints place strict requirements on the geometry of the
failures.

We have so far considered simple field geometries, and thus have not proved that plastic flow is the
generic macroscopic response of the crust. One can imagine situations in which mechanical failure in some region will cause magnetic-field reconfiguration that will in turn cause supercritical stress in another part of the crust, thus leading to a runaway process. 
This possibility has to be investigated further using specific magneto-elastic calculations.  
For now we remark that it is an open question
whether magnetar flares are associated with the crust failures; as was already explained in the introduction, viable alternative model is that
the flares are caused by reconnection events in the magnetosphere, in a manner similar to that observed
in the
solar flares (Lyutikov 2003, 2006).
Therefore, in our opinion the inferences about flare statistics based on phenomenological models for
the crust crystal failure (e.g., Perna \& Pons 2011) have to be made with caution.
 
We thank Andrei Beloborodov, Andrei Chugunov, Charles Horowitz, Rimma Lapovok, Bennett Link, and Jay
Melosh for stimulating discussions.

\section*{References}
\begin{footnotesize} \noindent
% Barat C. et al., 1983, A\& A, 126, 400\\
%Beloborodov, A.~in preparation\\
%Blandford, R., \& Thorne, K.~S., Application of Classical Physics, online course at \\
Ashby, M.~F., \& Sammis, C.~G., 1990, PAGEOPH, 133, 489\\
Blaes, O., Blandford, R., Goldreich, P., \& Madau, P., 1989, ApJ, 343, 839\\
Bridgeman, P.~W., 1945, Am.~J.~Sci., 243A, 90\\
Chamel, N., \& Haensel, P.~2008, Living Rev.~Relativity, 11, 10\\
Chugunov, A.~I., \& Horowitz, C.~J., 2010, MNRAS Letters, 407, 54\\
Gill, R., \& Heyl, J.~S., 2010, MNRAS, 407, 1926\\
Goldreich, P., \& Reisenegger, A., 1992, ApJ, 395, 250\\
%Haensel P., \& Potekhin A.~Y., 2004, A\&A, 428, 191\\
%http://www.ioffe.ru/astro/NSG/NSEOS/\\
%Haensel P., Potekhin A.~Y., Yakovlev D.~G., 2007. Neutron Stars 1: Equation of State
%and Structure (New York: Springer)\\
Horowitz C. J., Kadau K., 2009, Phys. Rev. Lett., 102, 191102\\
Hurley K., Boggs S. E., Smith D. M., Duncan R. C., Lin R.,
Zoglauer A., Krucker S., Hurford G., Hudson H., Wigger C., Hajdas W., Thompson C., Mitrofanov I., Sanin A., Boynton W., Fellows C., von Kienlin A., Lichti G., Rau A., 2005, Nature, 434, 1098\\
Hurley K., Cline T., Mazets E., Barthelmy S., Butterworth P., Marshall F., Palmer D., Aptekar R., Golenetskii S., Il'Inskii V., Frederiks D., McTiernan J., Gold R., Trombka J., 1999, Nature, 397, 41\\
Mazets E. P., Golentskii S. V., Ilinskii V. N., Aptekar R. L.,
Guryan I. A., 1979, Nature, 282, 587\\ 
%Israel G.~L. et al., 2005, ApJ, 628, L53\\
Jones, P.~B., 2003, ApJ, 595, 342\\
Kostrov, B.~V., 1966, J.~Appl.~Math.~Mech., 30, 1241\\
Landau, L.~D., \& Lifshits, E.~M.~1996, Elasticity theory, 
 (Pergamon press)\\
Levin, Y., \& van Hoven, M., 2011, MNRAS, 418, 659 \\
Link, B., Epstein, R.~I., 1996, ApJ, 457, 844\\
Lyutikov, M., 2003, MNRAS, 346, 540\\
Lyutikov, M., 2006, MNRAS, 367, 1594\\
Mereghetti S., 2008, Astron. Astrophys. Rev., 15, 225\\
Molinari, A., 1997, J.~Mech.~Phys.~Solids, 45, 1551\\
Palmer, D.M.,  \etal\ 2005, Nature,  434,  1107-1109\\
Perna, R., \& Pons, J.~A.,~2011, ApJ, 727, L51\\
Pons, J.~A., \& Geppert, U., 2007, A\&A, 470, 303\\
Steiner W., Watts A.~L., 2009, Phys.~Rev.~Letters, 103r1101S\\
%Strohmayer, T.~E. \& Watts, A.~L., 2005, ApJ, 632, L111\\
Thompson C., Duncan R. C., 1995, MNRAS, 275, 255 (TD95)\\
Thompson, C., Lyutikov, M., Kulkarni, S.~R., 2002, ApJ, 574, 332\\ 
%van Hoven, M., \& Levin, Y.~2011, MNRAS \\
%Watts, A.~L. \& Strohmayer T.~E., 2006, ApJ, 637, L117\\
Woods P. M., Thompson C., 2006, Soft gamma repeaters and anomalous X-ray pulsars: magnetar candidates. pp 547–586\\
\end{footnotesize}

%\end{multicols}

  \appendix

 \section{Normal modes in magnetised elastic medium with resistivity}
\label{Resistive}
In this Appendix we discuss the normal modes in the  elastic-resistive medium.
In Eqns (\ref{dyneq1}-\ref{dyneq2}), eliminating $B_x$ in favor of $\xi$, the equation for  the normal modes becomes
\be
\left( \partial_t - \eta \partial_z^2 \right) \left( \partial_t^2 -  c_{\rm el}^2 \partial_z^2 \right) \xi = v_A^2 \partial_t \partial_z^2 \xi 
\label{Eq}
\ee
This is a linear equation with constant coefficients and it can generally  be solved by separation of variable. Assuming $\xi = F(t) G(z)$, we find
\be 
{F^{(3)} \over F} - \left(c_t^2 {F' \over F} + \eta { F^{\prime \prime}  \over F} \right) {  G^{\prime \prime} \over G} - 
\eta c_{el}^2 {G^{(4)} \over G} =0
\label{self}
\ee
For  harmonic  spacial oscillations $\propto e^{-\om t + k z}$,  %$G^{\prime \prime} = -k^2 G$,  and assuming $F \propto e^{- i \om t}$, 
 the  dispersion equation is 
\be
\om^3 + i k^2 \eta \om^2 - k^2 c_t^2  \om -  i c_{el}^2 \eta k^4 =0 
\ee

In the limit $ \eta_{\rm res} \rightarrow 0$, the normal modes are
are 
\ba &&
\om^2_{1,2} =  c_t^2 k^2 \left( 1 \pm  { i k c_A^2 \eta_{\rm res} \over c_t^3} \right)
\nn &&
\om _3=- i {c_{el}^2 k^2 \over c_t^2} \eta
\ea
The modes \{1,2\} are just the resistivity-modified normal modes of the elastic medium with \Bf. 
They  become over-damped  (real part smaller than imaginary) for 
\be
k>  { c_t^3\over c_A^2 \eta_{\rm res} }
\ee
The mode 3 is the resistive mode.

\end{document}